# Anomalous binding of Fe atoms in chromium

S. M. Dubiel[*], J. Żukrowski and J. Cieślak

Faculty of Physics and Computer Science, AGH University of Science and Technology, al. A. Mickiewicza 30, 30-059 Kraków, Poland

**Abstract**

Binding of $^{57}$Fe atoms in a metallic chromium was investigated in a Cr-Fe alloy, containing less than 0.1 at% Fe enriched to ~95 % in $^{57}$Fe isotope, using $^{57}$Fe-site Mössbauer spectroscopy. The binding force was derived from the Debye temperature, $\theta_D$, that, in turn, was calculated from the temperature dependence of the central shift of the Mössbauer spectra recorded in the range of 80 to 330 K. Following a temperature dependence of the line width that shows a minimum at ~155 K, two temperature intervals were considered: a low temperature one (LT) ranging from 80 to 155 K, and the $\theta_D$ – value of 292 (12) K or 279 (34) K, and a high temperature one (HT) ranging from 155 to 330 K with the $\theta_D$ –value of 399 (15) or 399 (25) K, depending on the fitting procedure. The corresponding values of the harmonic force (spring) constant are: 35.4 N/m and 66.1 N/m or 33.8 N/m and 66.1 N/m for the LT and HT, respectively. This means that in the HT range the binding force of $^{57}$Fe atoms by the Cr matrix is by a factor of ~1.9 - 2 stronger than that in the LT range. This anomaly is possibly related with a different polarization of the spin-density waves in the LT and HT "phases".

[*] Corresponding author: dubiel@novell.ftj.agh.edu.pl (S. M. Dubiel)

Theoretically, charge-density waves (CDWs) and spin-density waves (SDWs) are connected with a broken-symmetry ground states. This has a close similarity to other broken-symmetry ground states of metals like superconductivity. In all these cases a gap, $2\cdot\Delta = 3.5\cdot k_B\cdot T_K$, develops in the dispersion relation curve at the Fermi wave vector, $k_F$ [1]. The name CDWs and/or SDWs follows from the harmonic modulation of the charge- and/or spin-density, respectively, with a periodicity of $\Lambda \sim 1/k_F$. The phenomenon of the SDWs, for which an electron-electron interaction is responsible, was originally introduced by Overhauser [2]. Meanwhile, it has been found in several systems including quasi-1D linear chain compounds like $TaS_3$ or $NbSe_3$, 2D layered transition metal dichalcogenides such as $TaS_2$, $CaI_2$ or $NbSe_2$ and 3D metals like $\omega$-Zr, Pb and Cr [3]. In metallic chromium, a subject of the present study, SDWs that originate from itinerant s- and d-like electrons exhibit interesting properties which are directly related to a density of electrons at the Fermi surface and its topology. From experimental viewpoint they can be investigated by a variety of experimental techniques [4,5], which enhances the attraction of the phenomenon. SDWs in chromium exist below the Néel temperature of $T_N \approx 313$ K [6,7] in two phases: (1) a high-temperature phase having a transverse polarization (TSDWs), and (2) a low-temperature phase that polarization is longitudinal (LSDWs). The temperature at which a transition between TSDWs and LSDWs takes place is called the spin-flip temperature, $T_{SF} \approx 123$ K. It should be mentioned, which is especially important in the contest of this study, that SDWs are extremely sensitive to various kinds of lattice imperfections like strain, defects and foreign atoms (impurities). The latter is of vital importance as far as certain experimental techniques such as perturbed angular correlations (PAC) and Mössbauer spectroscopy (MS) are concerned. None of them could have been used to study the issue unless probe nuclei have been introduced into the Cr-matrix. In these circumstances, a question arises whether or not the probe nuclei affect (pin) the original electronic structure, which is further complicated by the fact that in chromium CDWs with a half-periodicity coexist with the SDWs [4]. This means that impurities can potentially pin both SDWs and CDWs.

There are several papers that theoretically study the issue. Two types of impurities are considered: (1) nonmagnetic and (2) magnetic [1, 7-10]. A general conclusion from these calculations is that nonmagnetic impurities pin strongly CDWs and magnetic impurities do so for SDWs (strongly means that their effect is of the order of $\Delta$, which in the case of chromium is equal to ~0.1eV). However, there is not such agreement regarding the pinning of SDWs by nonmagnetic impurities. According to [8-10] the effect is weak i.e. much smaller than $\Delta$ which disagrees with the result found in [11], where the interaction between the nonmagnetic impurities and the SDWs can be as high as the value of $\Delta$. It should be added, that in the latter the SDW is treated as a superposition of two CDWs shifted in phase by $180^o$. It is also worth mentioning that according to some calculations magnetic impurities affect both LSDWs and TSDWs, while nonmagnetic ones can pin only LSDWs [10].

From experimental viewpoint, the situation seems to be very complicated, and in any case one cannot make a division into nonmagnetic and magnetic impurities [5]. Concerning the former, there are such ones that seem to act as ideal probe nuclei i.e. all features characteristic of the SDWs of a pure chromium can be revealed using them as probes. Here tantalum [12] and cadmium [13] in PAC experiments and tin in MS measurements [14-15] can be given as best examples. On the other hand, vanadium behaves like a SDWs killer as its presence quenches the SDWs decreasing $T_N$ at the rate of ~20 K/at%. Magnetic impurities have really strong effect. In particular, iron decreases $T_N$ and simultaneously changes the character of the SDWs from incommensurate to commensurate with a critical concentration of ~2.3 at%. Mössbauer effect measurements gave evidence that the $^{57}$Fe-sie hyperfine field is strongly reduced [17,18], in comparison with the corresponding quantity determined using the $^{119}$Sn-site effect [16]. However, interpretation of these results differs significantly. Whereas Wertheim concluded from his study that Fe atoms are not coupled to the host lattice, Herbert et al. arrived at an opposite opinion i.e. there is quite a strong interaction between the iron atoms and the SDW of the Cr matrix. They estimated the magnitude of the exchange field as 22.5 T. Furthermore, the temperature dependence of the effective hyperfine field was explained in terms of a spin-compensated state localized about the iron atoms (Kondo effect) and the Kondo temperature was estimated as $T_K = 60$ K. This value is, however, in a clear disaccord with the one found from a resistance minimum for a Cr-0.6at%Fe sample viz. ~140 K [19].

This unclear situation prompts additional investigations on the issue, all the more so, not all potential of the $^{57}$Fe-site MS has been in that respect explored in the past experiments. In particular, very important spectral parameter viz. the central shift, $CS$ has been completely neglected. As is well known, its temperature dependence can be expressed by the following equation:

$$\langle CS(T) \rangle = IS(T) + SOD(T) \qquad (1)$$

where the first term, $IS$, represents the isomer shift, a quantity related to a charge-density of s-like electrons at the nucleus site, and it was shown to weakly depend on temperature [20]. The second term, $SOD$, stays for the so-called second-order Doppler shift, and it is related to the atomic mean square displacement in the lattice, hence it is strongly temperature dependent, and can be easily calculated. This is why in practice one determines the Debye temperature, $\Theta_D$ assuming the total temperature dependence of $<CS>$ goes via $SOD$.

Neglecting then the temperature dependence of $IS$, and using the Debye model, the following equation can be written:

$$CS(T) = IS(0) - \frac{3k_B T}{2mc} \left[ \frac{3\Theta_D}{8T} + \left(\frac{T}{\Theta_D}\right)^3 \int_0^{\Theta_D/T} \frac{x^3}{e^x - 1} dx \right] \qquad (2)$$

where $m$ is the mass of $^{57}$Fe nucleus, $k_B$ is the Boltzmann constant and $c$ is the light velocity. A knowledge of $\Theta_D$-values can give us a hint on a binding strength, $\gamma$ of Fe atoms in the chromium lattice. There are few different approaches available. Following Visscher's simple-impurity theory for a simple-cubic lattice [21], the quantity $\Theta_{eff}$ i.e. the effective Debye temperature as determined in a Mössbauer experiment (e. g. from equ. (2)), is related to the Debye temperature of the matrix by:

$$\Theta_{eff} = (M_{Cr}/M_{Fe})^{1/2} (\gamma_{Fe-Cr}/\gamma_{Cr-Cr})^{1/2} \Theta_D \qquad (3)$$

where $M_{Cr}$ and $M_{Fe}$ are the masses of the host (Cr) and the impurity atom (Fe), while $\gamma_{Fe-Cr}$ and $\gamma_{Cr-Cr}$ are the spring constants of the impurity-host and the host-host binding. $\Theta_D$ is the Debye temperature of the host (Cr).

Gupta and Lal using a simple harmonic approximation derived the following formula for $\gamma$ [22]:

$$\gamma = \frac{M k_B^2 \Theta_D^2}{4 \hbar^2} \qquad (4)$$

which can be used to determine the spring constant itself, if the Debye temperature is known. The sample for the present investigation was prepared by melting elemental chromium (3N-purity) and iron enriched to ~95% in the $^{57}$Fe isotope mixed in such proportion to give ≤0.1 at% Fe. The melting process that was carried out in an arc furnace under a protective atmosphere of a pure argon was repeated 3 times to ensure a better homogeneity. For Mössbauer-effect measurements the ingot was filed with a diamond file and the powder was next annealed at 848 K for 24 hours in a quartz tube kept under a dynamic vacuum of ~2·10$^{-6}$ hPa to remove strain. $^{57}$Fe-site spectra, whose typical examples at different temperatures are shown in Fig. 1, were recorded in the temperature range of 80 – 330 K, both in cooling and heating cycles. As can be seen, down to 80 K the spectra have a form of a single line with a small broadening. For determining the Debye temperature the spectra were fitted with two different procedures: (I) each spectrum was treated as a single Lorentzian-shaped line, (II) based on a transmission integral method, each spectrum was considered as composed of two single-line components: one originating from the source and having a constant line width, and the other one originating from the sample and having a temperature dependent line width. Both methods yielded the *CS* and the line width, while the method II gave, in addition, a parameter proportional to the Lamb-Mössbauer factor. A temperature dependence of the half of the line width at half maximum, *HWHM*, as obtained with method I is shown in Fig. 2. As can be seen the dependence, which is reversible, show an anomalous behavior, and can be divided into two ranges: (i) a high temperature (HT) one for ~155 K ≤ T ≤ 330 K with a rather flat character, and (ii) a low temperature (LT) one for T ≤ ~ 155 K, where a steep increase of *HWHM* is observed. The border between the two ranges is marked by a shallow minimum. Such behavior agrees quite well with that found by Herbert et al [18]. Although its origin remains unknown, it seems worth mentioning that the temperature of ~155 K at which the minimum occurs is quite close to the Kondo temperature for the system of similar composition [19]. It also matches the temperature at which an anomaly in the SDW wave vector occurs for Fe-doped chromium [23].

Following the data presented in Fig. 2, *CS*-values were analyzed in terms of equation (2) in two separate ranges viz. HT and LT. The corresponding plots and the best fits of the experimental points obtained from the spectra fitting procedure I, are shown in Fig. 3. The values of $\Theta_D$ obtained by such procedure are 399±25 K for HT and 292±14 K for LT. Corresponding figures determined using the *CS*-values obtained by method II are displayed in Table 1 and they are in a good accord with the former. An unexpected behavior viz. a reversed dependence of $\Theta_D$ on temperature is worth mentioning. Normally, a lattice softens on increasing T i.e. $\Theta_D$ is smaller for higher T. In particular, for metallic Fe, it decreases from 430 K as found with the MS in the temperature range of 80 – 300 K, to 400 K for 300 K ≤ T ≤ 700 K and 310 K for 700 K ≤ T ≤ 1050 K [24,25]. At first glance it seems that the anomaly has to do with two different phases of the SDWs that exist in metallic chromium in the temperature domain i.e. transversely polarized ones (TSDW) that occur in the temperature range of 123 K to 313 K and longitudinally polarized ones (LSDW) that exist below 123 K. Here the division into the two temperature ranges HT and LT, from which $\Theta_D$-values were estimated, were made based on the anomaly in the *HWHM* behavior, which does not exactly coincides with the spin-flip temperature i.e. the temperature at which the TSDW-LSDW phase transition occurs. A steep increase in *HWHM*, hence in the spin-density, that starts at ~155 K is of interest in itself. One of a possible underlying reasons for the increase might be a spin-freezing mechanism known to occur in re-entrant spin-glasses [26]. However, an explanation of this aspect of the results presented is this letter needs more experimental investigation and theoretical calculations dedicated to the very issue. Here we want to express the anomaly in $\Theta_D$ in terms of the bonding strength, using for that purpose equations (3) and (4). In order to evaluate the spring constants, $\gamma$, we take into account the $\Theta_D$-values determined by means of the fitting procedures I and II and treat them in equations (3) and (4) as $\Theta_{eff}$. The relative and absolute values of $\gamma$ obtained for the investigated sample in such a way can be seen in Table 1.

It is clear that the value of the spring constant for the HT range is greater than the one for the LT range. The actual relationship between the two does not practically depend on the method of the spectra analysis used and it follows that the strength of bonding of $^{57}$Fe atoms in the HT range is by factor ~2 stronger than in the LT range. There is, of course, some isotope effect which has to be taken into account due to the fact that with the MS technique only $^{57}$Fe atoms were taken into account. To make a correction for it, we can calculate the relative strength of bonding of $^{57}$Fe atoms in metallic iron. Taking into account that the calorimetric Debye temperature for iron has a value of 463 K and that effective one as measured by the MS is 430 K, and using equ. (3) one arrives at $(\gamma^{57}_{Fe-Fe}/\gamma_{Fe-Fe}) = 0.865$. Referring our relative values of the spring constant, as found for the investigated sample – see Table 1, to the figure of 0.865 one arrives at a conclusion that the binding of $^{57}$Fe atoms in the iron matrix is about two times stronger than the binding of these atoms in the HT phase of chromium and about four times stronger than in the LT phase of chromium.

It is of interest to compare the presently found values of the relative spring constant with those found for $^{57}$Fe atoms in different metallic matrices. In particular, Steyert and Taylor using MS investigated $^{57}$Fe in Au, Cu, Ir, Pd, Pt, Rh and Ti and fund using the Vischers model that the relative spring constant was between 0.71 for Pt and 1.20 for Cu [27]. Sørensen and Trumpy investigated $^{57}$Fe dissolved in aluminium and arrived at the value of 0.51 [28]. The weak force between impurity and host atoms they connected with the very low solubility of Fe in the aluminium host. As the solubility of Fe in chromium host is high, while the impurity host forces are even weaker, we think that it is not the solubility that is responsible for the weakness of the bonding between Fe atoms and those of the host. It is rather the polarization of the SDWs that is responsible for the different binding force. However this issue cannot be solved experimentally, but rather theoretical calculations are needed for that. It is hoped the results presented in this paper will stimulate such calculations.

Table 1 Debye (effective) temperature, $\Theta_{eff}$ as determined with equation (2) for the high temperature (HT) and low temperature (LT) ranges. $\gamma$ stays for the spring constant. Relative $\gamma$ values as obtained with equation (3) for the HT and LT ranges are displayed in columns 4 and 5, while their absolute values as calculated from equation (4) can be seen in columns 6 and 7, respectively. The ratio, $R = \gamma_{Fe-Cr}(HT)/\gamma_{Fe-Cr}(LT)$ as calculated either from equation (3) or (4) is given in the last column.

| Fitting method | $\theta_{eff}$(HT) [K] | $\theta_{eff}$(LT) [K] | $\gamma_{Fe-Cr}/\gamma_{Cr-Cr}$ (HT) | $\gamma_{Fe-Cr}/\gamma_{Cr-Cr}$ (LT) | $\gamma_{Fe-Cr}$(HT) [N/m] | $\gamma_{Fe-Cr}$(LT) [N/m] | R |
|---|---|---|---|---|---|---|---|
| Single-line | 399 ± 15 | 292 ± 12 | 0.440 ± 0.037 | 0.235 ± 0.030 | 66.1 | 35.4 | 1.87 |
| Trans. Integ. | 399 ± 25 | 279 ± 34 | 0.440 ± 0.061 | 0.225 ± 0.083 | 66.1 | 33.8 | 1.96 |

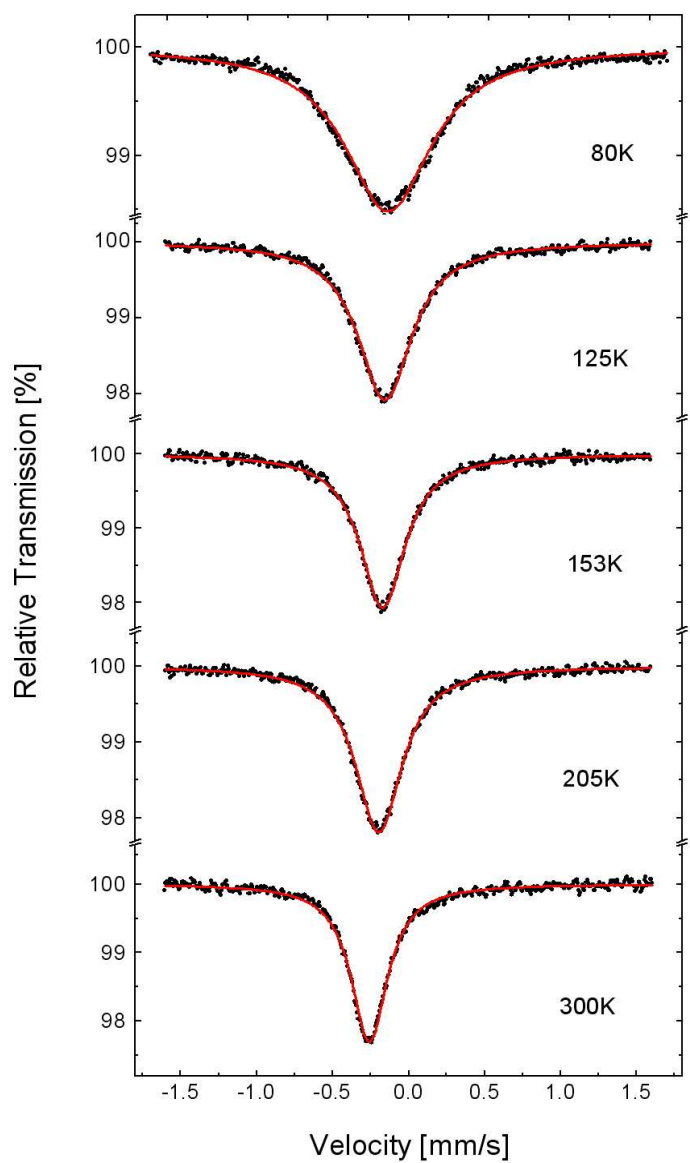

Fig. 1 $^{57}$Fe site Mössbauer spectra measured at various temperatures shown. Solid lines are the best fits to experimental data with procedure I.

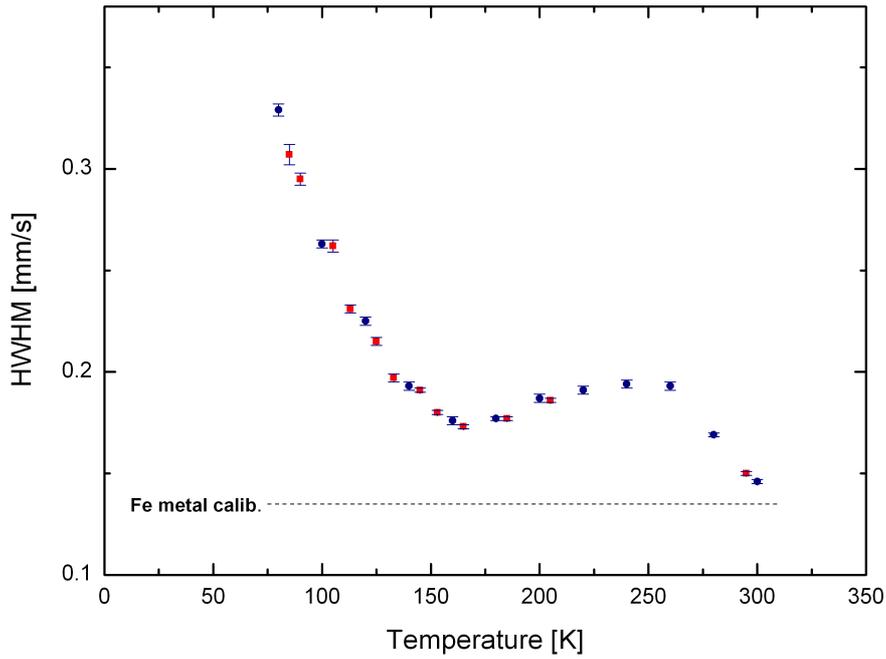

Fig. 2 Color on line: Half width at half maximum, *HWHM*, as a function of temperature, *T*, as evaluated from the Mössbauer spectra with procedure I. The corresponding quantity for the calibration metallic-Fe foil is marked by a horizontal dashed line. In blue are indicated points measured on increasing *T,* and those in red on decreasing *T*.

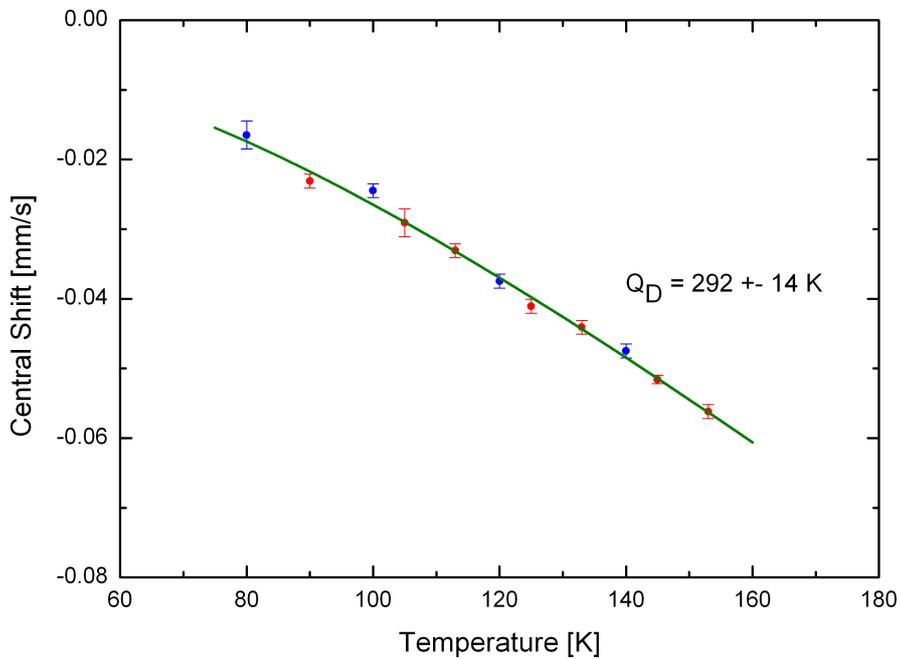

Fig. 3a Temperature dependence of the central shift, as determined with procedure I for the LT range. The solid line stays for the fit with eq. 2). The data points in blue were recorded on increasing T, while those in red on decreasing T.

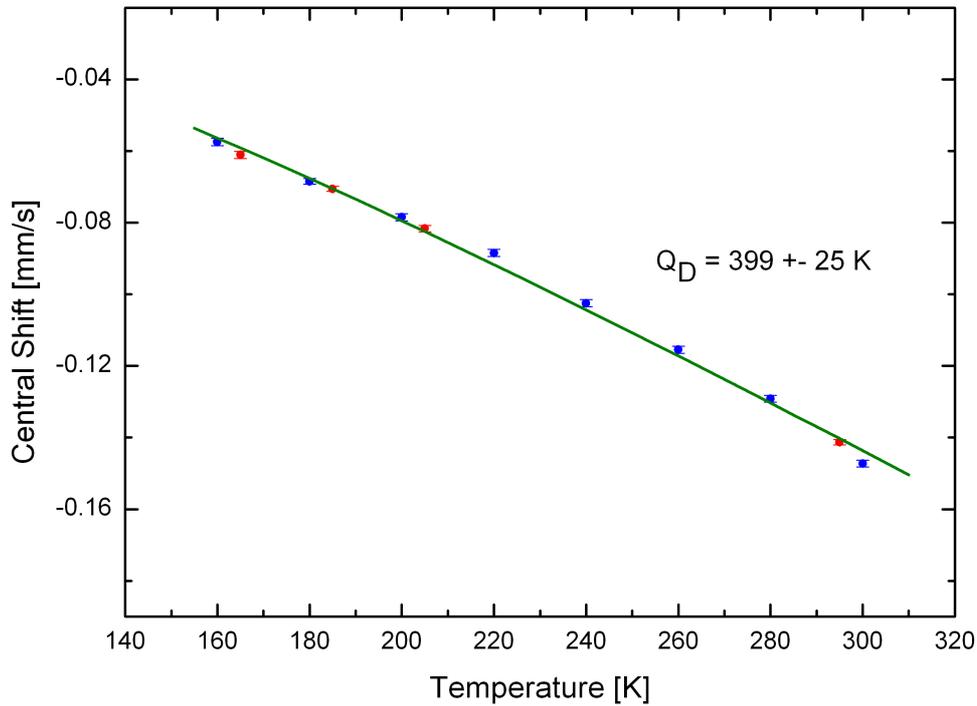

Fig. 3b Temperature dependence of the central shift, as determined with procedure I for the HT range. The solid line stays for the fit with eq. 2). The data points in blue were recorded on increasing T, while those in red on decreasing T.